# Multimodal Machine Learning for Integrating Heterogeneous Analytical Systems

Shun Muroga[1,2]*, Hideaki Nakajima[1], Taiyo Shimizu[1], Kazufumi Kobashi[1], Kenji Hata[1,2]

1: Nano Carbon Material Research Institute, National Institute of Advanced Industrial Science and Technology (AIST), Tsukuba Central 5, 1-1-1, Higashi, Tsukuba, Ibaraki, 305-8565, Japan

2: Materials DX Research Center, National Institute of Advanced Industrial Science and Technology (AIST), Tsukuba Central 5, 1-1-1, Higashi, Tsukuba, Ibaraki, 305-8565, Japan

*Corresponding author. Shun Muroga, E-mail: muroga-sh@aist.go.jp

**Abstract**

Understanding structure–property relationships in complex materials requires integrating complementary measurements across multiple length scales. Here we propose an interpretable "multimodal" machine learning framework that unifies heterogeneous analytical systems for end-to-end characterization, demonstrated on carbon nanotube (CNT) films whose properties are highly sensitive to microstructural variations. Quantitative morphology descriptors are extracted from SEM images via binarization, skeletonization, and network analysis, capturing curvature, orientation, intersection density, and void geometry. These SEM-derived features are fused with Raman indicators of crystallinity/defect states, specific surface area from gas adsorption, and electrical surface resistivity. Multi-dimensional visualization using radar plots and UMAP reveals clear clustering of CNT films according to crystallinity and entanglements. Regression models trained on the multimodal feature set show that nonlinear approaches, particularly XGBoost, achieve the best predictive accuracy under leave-one-out cross-validation. Feature-importance analysis further provides physically meaningful interpretations: surface resistivity is primarily governed by junction-to-junction transport length scales, crystallinity/defect-related metrics, and network connectivity, whereas specific surface area is dominated by intersection density and void size. The proposed multimodal machine learning framework offers a general strategy for data-driven, explainable characterization of complex materials.

**Keywords**:
Multimodal machine learning, Explainable machine learning, Scanning electron microscopy, Carbon nanotubes

**Introduction**

The functional performance of materials is governed not only by chemical composition but also by structural features spanning hierarchical length scales from the nanoscale to the macroscale. Relevant factors include crystallinity, defects, interfaces, phase separation, orientation, inter-material entanglement, and pore architecture. Because any single analytical technique probes only a portion of this complexity, a single measurement rarely provides a comprehensive picture. This constraint impedes identification of the dominant determinants of mechanical, thermal, and electrical properties, complicates interpretation of the underlying phenomena, and ultimately hinders the formulation of practical design guidelines for both materials and processing conditions. Therefore, elucidating

structure–property relationships in complex materials requires multifaceted characterization in which complementary analytical methods are combined and interpreted across multiple observational scales.

Alongside advances in measurement, data science has become indispensable for extracting knowledge from multivariate datasets obtained across multiple specimens. In analytical chemistry, multivariate analysis has a long history, exemplified by chemometrics [1–4], which has been actively developed worldwide. For spectral data in particular, indirect property estimation and anomaly detection - conceptually analogous to soft sensing - have been widely implemented, with partial least squares (PLS) regression [5–7] serving as a standard approach. This is largely enabled by systematic spectral variations consistent with the Lambert–Beer law and by strong collinearity of absorbance changes across wavelengths. In addition, methods such as two-dimensional correlation spectroscopy [8,9] have been developed to enhance subtle spectral variations by introducing an additional analytical "viewpoint," thereby facilitating extraction of underlying fluctuations that may be difficult to detect by inspection alone.

Applying these concepts to diverse and complementary measurement datasets is essential for understanding complex materials. In this context, multimodal machine learning is expected to play an increasingly important role in analytical chemistry. Here, "multimodal" denotes the integration of heterogeneous data modalities, analogous to how humans combine information from multiple sensory inputs to make decisions. Although multimodal approaches were initially advanced in areas such as emotion analysis and medical diagnosis, they are now rapidly expanding into materials science. We have investigated multimodal machine learning methodologies that integrate distinct measurement datasets, extract latent material features, and accelerate both scientific discovery and optimization of materials and manufacturing processes. Specifically, we have developed frameworks for property prediction and inverse design by combining multiple deep generative models with unified deep learning architectures that integrate measurement data capturing different physicochemical structures [10]. We have also reported methods to infer the temporal or ordinal progression of structural evolution from complementary multimodal measurements [11], incorporate descriptors of solvent chemical speciation [12], and represent particle size distributions as quantitative features to clarify their relationships with bulk material performance [13].

In the present paper, we focus on carbon nanotube (CNT) films as a model system and present a multimodal machine learning framework that processes integrated measurement datasets to extract interpretable features of complex structures. CNTs are cylindrical carbon-based materials with exceptional properties [14], including strength approximately twenty times that of steel, thermal conductivity roughly ten times that of copper, density about half that of aluminum, and electron mobility on the order of ten times that of silicon. These characteristics have enabled a broad range of applications, from laboratory demonstrations to commercial products, including electronic components such as transparent conductive films [12,15], thin film transistor [16–20], thermoelectric devices [21,22], capacitors [23,24], conductive fibers [13,25], and polymer-based composites such as rubbers [26–31], thermosetting resins [10,32–34], and thermoplastic resins [35–38]. At the same time, CNTs remain challenging in terms of material and process control because their hierarchical structures [24,39,40] span multiple length scales: component-level properties are influenced by factors ranging from atomic-scale defect structures to meso- and macroscale features such as porosity and CNT entanglement.

In this study, we perform CNT feature analysis using multimodal machine learning by linking complementary datasets obtained from scanning electron microscopy, Raman spectroscopy, gas adsorption measurements, and electrical conductivity measurements (Fig. 1). To improve the interpretability of CNT film microstructure, we derive topological descriptors from film morphology and integrate them with features from other modalities within a unified learning framework, aiming to elucidate their relationships with electrical surface resistivity and specific surface area.

## Materials and methods

### Fabrication of CNT Films

In this study, seven commercially available carbon nanotubes (CNTs) were used: four multi-walled CNTs (FloTube 9000, NC7000, K-nanos 100p, and JC142) and three single-walled CNTs (SG-CNT HT, TUBALL, Meijo eDIPS EC2.0). These materials were purchased from or provided by JiangSu CNano Technology Co., Ltd., Nanocyl SA, Kumho Petrochemical Co., Ltd., JEIO Co., Ltd., Zeon Corporation, OCSiAl, and Meijo Nano Carbon Co., Ltd. Hereafter, the CNTs are referred to by the following abbreviations: Cnano, Nanocyl, Knano, JEIO, SG, Tuball, eDIPS. Each CNT was dispersed in methyl isobutyl ketone using a bead-milling process. The resulting dispersions were filtered through a PTFE membrane filter and then dried under vacuum to obtain CNT films. Detailed information on the CNT films has been reported in our previous work [24].

### Characterization

Scanning electron microscopy (SEM) observations were conducted using a field-emission SEM (SU8220, Hitachi) operated at an accelerating voltage of 5 kV and an emission current of 10 μA. For each CNT film, more than ten SEM images were acquired to ensure a representative evaluation of the surface morphology. Raman spectroscopy was performed using a confocal Raman microscope (inVia, Renishaw) with a 532 nm excitation laser. The G/D intensity ratio, which reflects the degree of structural defects associated with $sp^2$- and $sp^3$-bonded carbon in CNTs, was calculated as an indicator of crystallinity [41]. Electrical surface resistivity was measured using a four-point probe method with a resistivity meter (Loresta-GP MCP-T610, Mitsubishi Chemical Analytech). The Brunauer–Emmett–Teller (BET) specific surface area was determined from nitrogen adsorption isotherms [39] measured at 77 K using pore size distribution analyzers (BELSORP-mini and BELSORP-max, MicrotracBEL).

### Data Analysis

SEM images were preprocessed to extract topological features of the CNT films using MATLAB R2020b with the Image Processing Toolbox. Multimodal machine learning on datasets obtained from heterogeneous analytical systems was performed using Python (version 3.11.13) with several modules, including NumPy (version 1.26.4), pandas (version 2.1.4), and scikit-learn (version 1.4.2).

## Results and discussion

As a first step, we describe the SEM image analysis workflow used to translate CNT-film microstructures into quantitative structural features (Fig. 2). As illustrated in Fig. 2 (a), raw SEM images are first binarized to separate the CNT phase (white) from the void phase (black). The binarized images are then skeletonized to extract network centerlines, and intersection points are detected. This procedure enables consistent computation of descriptors such as phase fractions (CNT/void), skeleton-based curvature and orientation (reflecting anisotropy), and intersection density (reflecting entanglement). Representative binarized images for different CNT films are shown in Fig. 2b, demonstrating that qualitative differences - such as bundle density, alignment, and junction frequency - can be mapped onto comparable numerical features. These image-derived descriptors are subsequently compared across samples to quantitatively discuss structural variations in CNT networks. Previous studies have investigated network structures primarily through theoretical frameworks such as effective medium theory and percolation theory, which have been extensively applied to rod-shaped metallic nanowire networks [42–44]. In contrast, for CNT films, experimental quantification of network morphology has been reported mainly for thin, sparse networks used in CNT transistors [16,18,20]. Here, we demonstrate that structural features can be automatically and reliably extracted from SEM images and, when integrated with complementary measurements via multimodal machine learning, enable quantitative comparison of CNT network morphologies across materials and assessment of their contributions to macroscopic properties. This approach provides a practical route to data-driven interpretation of complex structures across a wide range of material systems.

We then visualize structural features obtained from heterogeneous analytical systems - SEM-derived network descriptors and Raman spectroscopy - across diverse CNT films. As summarized in Table 1, the SEM-based descriptors span broad ranges, including mean CNT diameter (11.0–15.7 nm), mean curvature (0.083–0.130), mean CNT length between intersections (4.86–6.67), intersection density (0.00327–0.00558), void area ratio (0.424–0.440), and orientation functions (0.597–0.647 for CNTs and 0.241–0.260 for voids). Raman- and property-related metrics also vary substantially, including the G/D ratio (0.61–37.9), specific surface area (220–1050 $m^2$/g), and surface resistivity (0.58–35.4 Ω/□). Based on these multidimensional descriptors, Fig. 3 (a) presents radar-plot "structural fingerprints" that highlight dominant characteristics, such as markedly elevated G/D ratios for Tuball and eDIPS, and high intersection density with small void diameter for SG and JEIO. To further visualize overall similarities and differences in the multi-dimensional space, we applied Uniform Manifold Approximation and Projection (UMAP)[45], a manifold-learning method for dimensionality reduction. The two-dimensional UMAP embedding (Fig. 3(b)) organizes the CNT films primarily into three groups: (1) high-crystallinity SWCNTs (Tuball and eDIPS), (2) high–specific-surface-area CNTs (SG and JEIO), and (3) the remaining MWCNTs (Cnano, Nanocyl, and Knano). This clustering suggests that the computed structural features capture rich information on complex CNT architectures that is relevant for comparing film properties.

Fig. 4 provides data-driven evidence that CNT-film properties are governed not by a single descriptor but by the combined contributions of network architecture (entanglement, transport length scales, and void geometry) and crystallinity/defect states. The correlation map in Fig. 4 (a) indicates multivariate dependencies of surface resistivity and specific surface area on multiple structural features, motivating an interpretation that accounts for interacting morphological factors (e.g., entanglement–alignment coupling and void–bundling effects). We therefore trained regression models using multimodal structural features and identified dominant contributors via permutation feature importance under leave-one-out cross validation (LOOCV). For surface resistivity, the mean CNT length between intersections ranks highest, followed by the Raman G/D ratio and intersection density (Fig. 4 (b)). This ranking is consistent with a physical picture in which electrical transport is influenced by the effective path length set by junction-to-junction spacing, the percolation and contact-resistance statistics governed by junction density, and scattering associated with crystallinity/defects. For specific surface area, intersection density and mean void diameter dominate (Fig. 4 (c)), highlighting the roles of fine-scale connectivity and pore-size architecture in determining accessible surface area. Performance benchmarking under LOOCV further shows that the nonlinear eXtreme Gradient Boosting (XGBoost) regressor [46] achieves the best accuracy among the tested linear and nonlinear methods (Table 2). Together, these results support the presence of nonlinear coupling among junction-scale connectivity, pore geometry, and crystallinity, while the importance analysis provides an interpretable route to identify physically meaningful structure–property relationships through multimodal machine learning.

**Conclusion**

In this work, we developed a multimodal machine learning framework that integrates information from heterogeneous analytical systems - including scanning electron microscopy, Raman spectroscopy, gas adsorption tests, and electrical conductivity tests - and demonstrated an end-to-end workflow for CNT films, spanning structural feature extraction and visualization through property prediction and interpretation. By fusing SEM-derived morphological descriptors with Raman-based indicators of crystallinity and defect states, we quantitatively captured structural variability across films and, using a nonlinear regression model (XGBoost) together with feature-importance analysis, identified physically meaningful structure–property relationships. Our results indicate that surface resistivity and specific surface area are not governed by any single descriptor but instead emerge from coupled effects of network connectivity, junction-to-junction transport length scales, void-size architecture, and defect-related characteristics.

Beyond CNT films, these findings underscore a broader implication for complex materials characterization: interpretation must move beyond parallel, instrument-by-instrument analyses toward an integrated and explainable linkage between distributed multimodal measurements and target properties. In this regard, multimodal machine learning provides a practical route to unify rapidly expanding heterogeneous analytical systems and to enable quantitative, interpretable evaluation of complex materials, with strong potential to serve as a standard analytical foundation across diverse materials platforms.

**Acknowledgements**: This work was partially supported by Japan Society for the Promotion of Science (JSPS) KAKENHI (No. JP22K14571).

**Author Contributions**
Conceptualization: S.M., Methodology: S.M., Software: S.M., Formal Analysis: S.M., Investigation: S.M., H.N., T.S. K.K. K.H., Visualization: S.M., Writing – Original Draft: S.M.

**Data availability:** Additional supporting data generated during the present study are available from the corresponding author upon reasonable request.

**Declarations**
**Conflicts of interest:** The author declares having no conflicts of interest or competing interests.

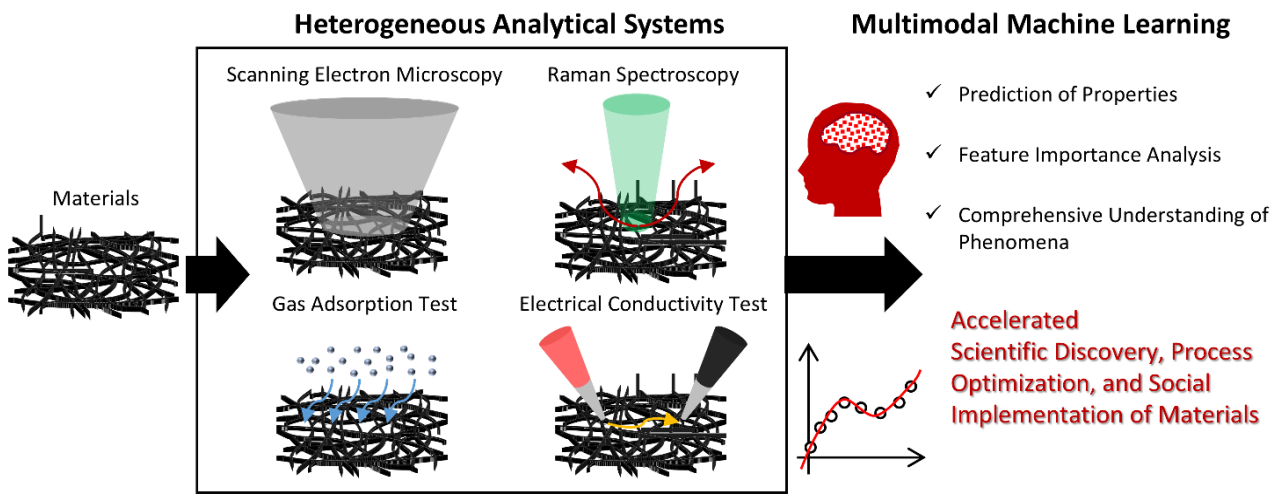

**Fig. 1.** Schematic illustration of a multimodal machine learning framework that integrates heterogeneous analytical techniques for comprehensive materials characterization.

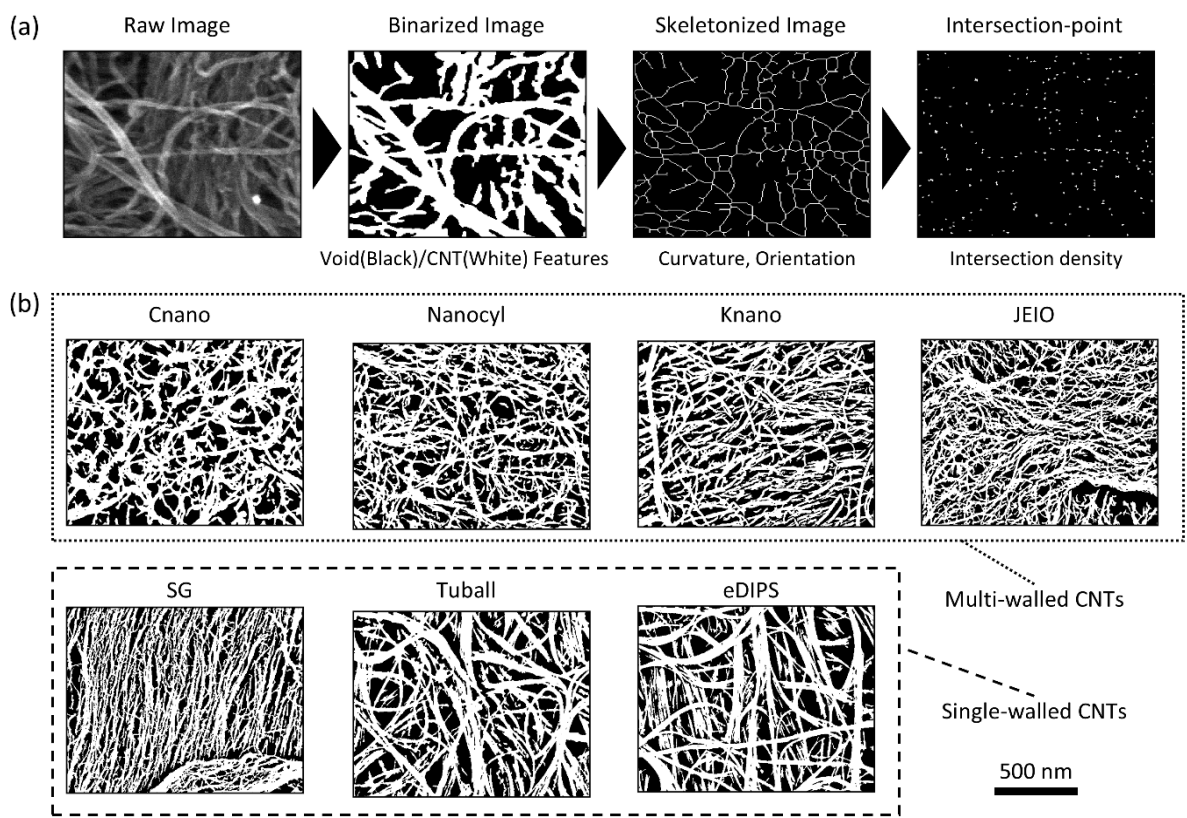

**Fig. 2.** Feature extraction from scanning electron microscope (SEM) images. (a) Image processing workflow. (b) Representative binarized SEM images of different types of CNT films.

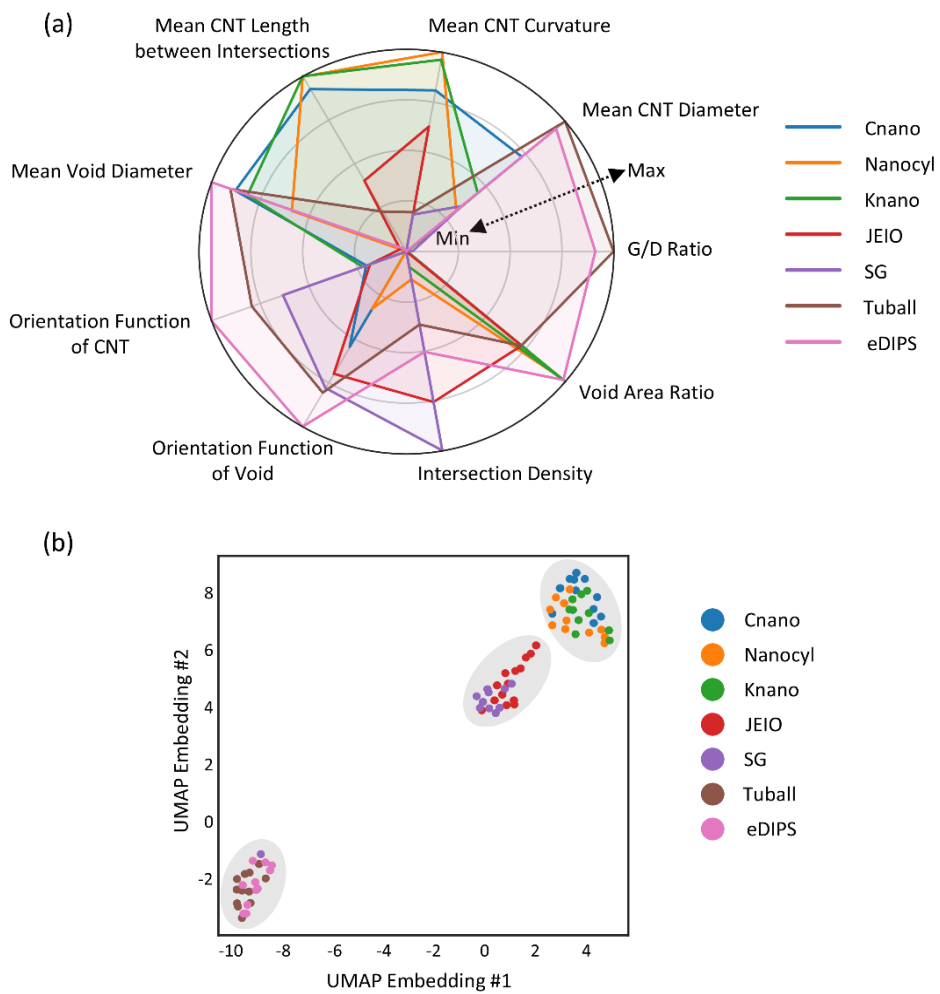

**Fig. 3.** Visualization of structural features of materials. (a) Radar plots of structural features of CNT films derived from SEM images and Raman spectroscopy. (b) Two-dimensional embedding of structural features using Uniform Manifold Approximation and Projection (UMAP).

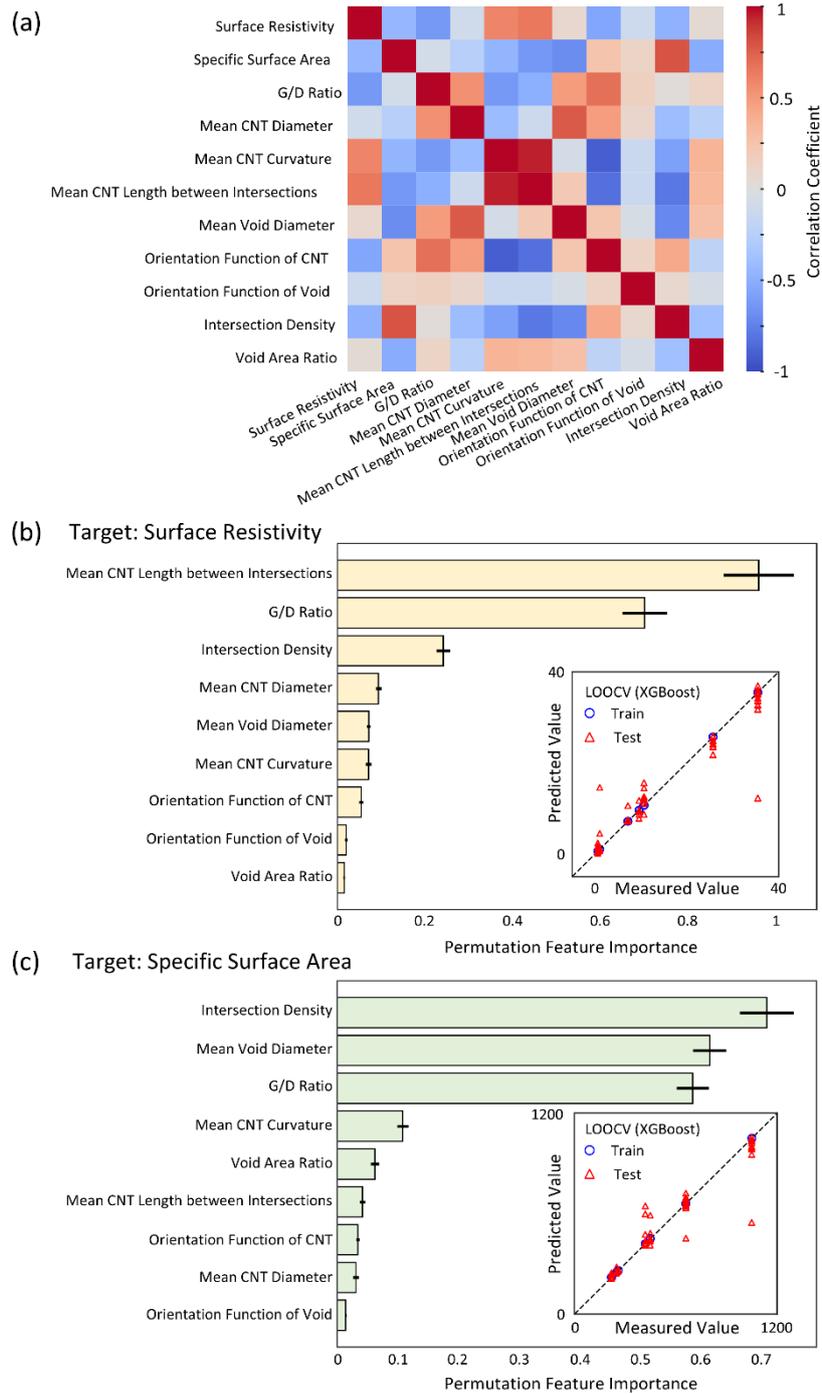

**Fig. 4.** Multimodal machine learning analysis of structural features and material properties. (a) Correlation map between structural features and properties. (b, c) Feature importance in regression models for (b) surface resistivity and (c) specific surface area of CNT films.

**Table 1.** Structural features and properties of CNT films derived from heterogeneous analytical systems for multimodal machine learning.

| Name | Type | Mean CNT Diameter [nm] | Mean CNT Curvature | Mean CNT Length between Intersections [nm] | Mean Void Diameter [nm] | Orientation Function of CNT | Orientation Function of Void | Intersection Density | Void Area Ratio | G/D Ratio | Surface Resistivity [Ω/□] | Specific Surface Area [m²/g] |
|---|---|---|---|---|---|---|---|---|---|---|---|---|
| Cnano | Multi-walled CNT | 14.5 | 0.121 | 6.54 | 32.0 | 0.607 | 0.251 | 0.00327 | 0.436 | 0.73 | 35.4 | 220 |
| Nanocyl | Multi-walled CNT | 12.5 | 0.130 | 6.67 | 29.5 | 0.597 | 0.247 | 0.00360 | 0.439 | 0.66 | 25.7 | 260 |
| Knano | Multi-walled CNT | 13.1 | 0.128 | 6.67 | 31.5 | 0.608 | 0.241 | 0.00345 | 0.440 | 0.72 | 10.7 | 250 |
| JEIO | Multi-walled CNT | 11.0 | 0.113 | 5.59 | 24.7 | 0.607 | 0.254 | 0.00502 | 0.436 | 0.61 | 7.16 | 660 |
| SG | Single-walled CNT | 12.7 | 0.092 | 4.86 | 24.4 | 0.629 | 0.256 | 0.00558 | 0.424 | 1.70 | 9.57 | 1050 |
| Tuball | Single-walled CNT | 15.7 | 0.092 | 5.28 | 32.3 | 0.637 | 0.257 | 0.00412 | 0.436 | 37.9 | 1.00 | 450 |
| eDIPS | Single-walled CNT | 15.5 | 0.083 | 4.89 | 33.1 | 0.647 | 0.260 | 0.00444 | 0.440 | 34.6 | 0.58 | 420 |

**Table 2.** Performance comparison of regression models within a multimodal machine learning framework for predicting CNT film properties.

| Target | Method | LOOCV Test RMSE | LOOCV Test MAE | LOOCV Test $R^2$ |
|---|---|---|---|---|
| Surface Resistivity | eXtreme Gradient Boosting (XGBoost) Regressor | 3.31 | 1.33 | 0.925 |
| Surface Resistivity | Random Forest Regressor | 4.80 | 3.18 | 0.843 |
| Surface Resistivity | Support Vector Machine Regressor (RBF kernel) | 6.19 | 4.31 | 0.740 |
| Surface Resistivity | Gaussian Process Regressor (Matérn kernel) | 6.57 | 5.14 | 0.706 |
| Surface Resistivity | Least Absolute Shrinkage and Selection Operator (LASSO) | 8.29 | 6.81 | 0.532 |
| Surface Resistivity | Partial Least Squares (PLS) Regressor | 8.68 | 7.03 | 0.487 |
| Specific Surface Area | eXtreme Gradient Boosting (XGBoost) Regressor | 73.7 | 28.3 | 0.926 |
| Specific Surface Area | Random Forest Regressor | 98.0 | 52.1 | 0.869 |
| Specific Surface Area | Least Absolute Shrinkage and Selection Operator (LASSO) | 133.6 | 98.0 | 0.757 |
| Specific Surface Area | Gaussian Process Regressor (Matérn kernel) | 150.1 | 79.5 | 0.693 |
| Specific Surface Area | Partial Least Squares (PLS) Regressor | 150.6 | 113.0 | 0.691 |
| Specific Surface Area | Support Vector Machine Regressor (RBF kernel) | 154.1 | 91.8 | 0.677 |